\documentclass[prb,aps,floats,superscriptaddress,nofootinbib,twocolumn,floatfix]{revtex4}
\usepackage{exscale}                  
\usepackage[intlimits]{amsmath}       
\usepackage{amsfonts}
\usepackage{amssymb,amscd}
\usepackage{wrapfig}
\usepackage{graphicx}
\usepackage{array}
\usepackage{bbm}
\usepackage{multirow}
\usepackage[T1]{fontenc}
\usepackage{times}
\usepackage{subfigure}
\usepackage[x11names]{xcolor}

\newcommand{\nf}{N_{f}\,}
\newcommand{\vf}{v_{F}\,}
\newcommand{\vd}{v_{\Delta}\,}
\newcommand{\qed}{$\mbox{QED}_3$}

\newcommand{\beq}{\begin{equation}}
\newcommand{\eeq}{\end{equation}}
\newcommand{\Eq}[1]{Eq.~(\ref{#1})}

\begin{document}

\title{Effects of Anisotropy in $\mbox{QED}_3$ from Dyson-Schwinger equations in a box}

\author{Jacqueline~A.~Bonnet}
\affiliation{Institut f\"ur Theoretische Physik, 
 Universit\"at Giessen, 35392 Giessen, Germany}
\author{Christian~S.~Fischer}
\affiliation{Institut f\"ur Theoretische Physik, 
 Universit\"at Giessen, 35392 Giessen, Germany}
\affiliation{Gesellschaft f\"ur Schwerionenforschung mbH, 
  Planckstr. 1  D-64291 Darmstadt, Germany.}
\author{Richard Williams}
\affiliation{Institut f\"ur Physik, Karl-Franzens Universit\"at Graz, 
Universit\"atsplatz 5, A-8010 Graz, Austria.}

\begin{abstract}
We investigate the effect of anisotropies in the fermion velocities
of 2+1 dimensional QED on the critical number $N_f^c$ of fermions for 
dynamical mass generation. Our framework are the Dyson-Schwinger 
equations for the gauge boson and fermion propagators formulated in a 
finite volume. In contrast to previous Dyson-Schwinger studies we do 
not rely on an expansion in small anisotropies but keep the full velocity 
dependence of fermion equations intact. As result we find sizable 
variations of $N_f^c$ away from the isotropic point in agreement with 
other approaches. We discuss the relevance of our findings for models 
of high-$T_c$ superconductors. 
\end{abstract}

\maketitle

\section{Introduction \label{sec:introduction}}

Quantum electrodynamics in two space and one time dimensions (QED$_3$) has been
studied intensely in the past. QED$_3$ is a comparably simple 
quantum field theory which nevertheless displays a rich structure of nonperturbative 
phenomena such as (a variant of) confinement and dynamical chiral symmetry breaking 
plus the associated generation of a fermion mass. In condensed matter physics, 
QED$_3$ and variants thereof have been suggested as effective theories to 
describe strongly interacting systems such as high temperature  
superconductors \cite{Franz:2002qy,Herbut:2002yq} or graphene 
\cite{Novoselov:2005kj,Gusynin:2007ix}. 

In the context of high-temperature superconductivity the chirally broken and
symmetric phases of QED$_3$ are associated with the antiferromagnetic spin-density
wave state (SDW) and the pseudogap phase of the cuprate compound
described as `algebraic 
Fermi liquid' \cite{Franz:2002qy,Herbut:2002yq}. Here, the fermions
of QED$_3$ represent chargeless quasi-particles, which are low-energy excitations 
close to the nodes of the (d-wave symmetric) gap-function. These spinons interact 
with collective, topological excitations of the gap that can be described
by a $U(1)$ gauge theory. Furthermore, since the motion of the quasi-particles 
is confined to the two-dimensional copper-oxide planes in these systems 
one ends up with quantum electrodynamics in (2+1) dimensions. 

There is a long history of debate on the critical number $N_f^c$ of fermions,
where the isotropic version of QED$_3$ undergoes the transition from the
chirally broken into the chirally symmetric phase. At zero temperature, 
if $N_f^c > 2$ then underdoping drives the superconductor directly into the 
SDW phase, while if $N_f^c < 2$ the system will first go through the  
pseudogap phase. Indeed, in the continuum studies mostly using Dyson-Schwinger 
equations \cite{Franz:2002qy,Appelquist:1988sr,Nash:1989xx,Maris:1996zg,Fischer:2004nq} 
this number has been found in the range of $N_f^c=3.5-4$ using Ward-identity 
improved fermion-photon vertices \cite{Fischer:2004nq}. This range seems to 
include the critical number extracted from lattice gauge theory 
\cite{Hands:2002dv,Hands:2004bh,Strouthos:2008hs}, provided the (large) 
corrections due to the finite lattice volume have been taken into 
account \cite{Gusynin:2003ww,Goecke:2008zh}. 

When comparing with the experimental situation, however, isotropic QED$_3$
may not be the appropriate choice. In fact there are sizable anisotropies in
the cuprate compounds due to large differences in the Fermi velocity 
$v_F$ and a second velocity $v_D$ related to the amplitude of the 
superconducting order parameter. This fermionic anisotropy can be as large as 
\cite{Chiao:2000} $\lambda = \vf / \vd \sim 10$. Furthermore, both velocities
are also different from the speed of light, $c_s$, which is the third scale
in the theory. Thus in the effective theory all three Euclidean axes have to 
be distinguished and it is an important question how and whether the critical
number of fermion flavors $N_f^c$ depends on $(c_s,\vf,\vd)$. In the literature, 
this situation has begun to be explored in a number of approaches. First studies
using renormalisation group and Dyson-Schwinger equations 
\cite{Franz:2002qy,Lee:2002qza} used an expansion only valid for small anisotropies. 
There it has been concluded that anisotropy is an irrelevant phenomenon in
the renormalisation group sense and does not lead to drastic changes in $N_f^c$.
In contrast, lattice gauge theory \cite{Hands:2004ex,Thomas:2006bj} and a 
recent study of the strength of the fermion-photon coupling \cite{Concha:2009zj}
provided evidence for the opposite picture: according to these works anisotropy 
is relevant and its increase leads to a decrease of the critical number of 
fermion flavors $N_f^c$. This opens up the possibility that $N_f^c$ may go
down from $N_f^c=3.5-4$ in the isotropic case to $N_f^c < 2$ for values of 
anisotropy relevant for the high-T$_c$ superconducting systems. However, further
studies are necessary to pin down quantitative predictions for this effect.

In this work we reexamine the situation of anisotropic QED$_3$ from the perspective
of Dyson-Schwinger equations. We go beyond the small anisotropy expansions used
in Refs.~\cite{Franz:2002qy,Lee:2002qza} and solve the Dyson-Schwinger
equations for the nonperturbative, anisotropic fermion propagator numerically 
using on the one hand the $1/N_f$-approximation and on the other hand an improved
approximation of the photon propagator as input. These technical details 
are summarized in section \ref{sec:technical}. Our results, presented in section
\ref{results}, confirm the findings from the lattice \cite{Hands:2004ex,Thomas:2006bj}
and from the fermion-photon coupling strength \cite{Concha:2009zj}. We discuss 
qualitative aspects of the dependence of $N_f^c$ on the anisotropies and identify
meaningful approximation schemes for the DSEs. We conclude
with a summary in section \ref{summary}.

\section{technical details \label{sec:technical}}

\subsection{The Dyson--Schwinger equations in anisotropic $\mbox{QED}_3$ 
\label{subsec:anisotropicdses}}

The details of the formulation of anisotropic QED$_3$ have been discussed in
Refs.~\onlinecite{Franz:2002qy,Lee:2002qza,Hands:2004ex}. Here we only summarize the most 
important elements using the notation of Ref.~\onlinecite{Franz:2002qy}.
The fermions are represented in the four component spinor representation 
obeying the Clifford algebra
$\lbrace \gamma_{\mu},\gamma_{\nu}\rbrace=2\,\delta_{\mu\nu}$.
We consider $\nf$ fermion flavors and the anisotropic fermionic velocities 
$\vf$ and $\vd$ which will be included in the metric-like factor $g_{ i,\mu\nu}$. 
In the high-temperature superconducting (HTS) system, the d-wave symmetry of the 
gap function on the Fermi surface leads to four nodes, which can be grouped
into two distinct pairs \cite{Herbut:2002yq}. This is accounted for by the 
index $i=1,2$. The ``\emph{metric}'' is given by:

\begin{equation}
\left(g_{1}^{\mu\nu}\right)=\left(\begin{array}{ccc}
                                        1 &    0     & 0      \\
                                        0 & \,\vf^{2}& 0      \\
                                        0 &    0     & \vd^{2}
                                  \end{array}
                            \right)
\label{eq:metric 1}
\end{equation}
 and 
\begin{equation}
\left(g_{2}^{\mu\nu}\right)=\left(\begin{array}{ccc}
                                         1&    0      &  0    \\
                                         0&  \vd^{2}  &  0    \\
                                         0&   0       &\vf^{2}
                                  \end{array}
                            \right).
\label{eq:metric 2}
\end{equation}\\

\begin{figure}[b]
		\begin{center}
			\subfigure{\label{fig:fermiondse}
                                  \includegraphics[width=0.9\columnwidth]{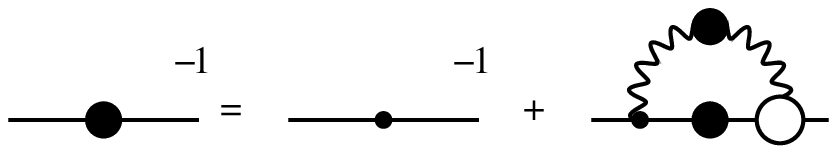}}
	                           \hspace{0.08\columnwidth}\vspace*{3mm}\\
	                \subfigure{\label{fig:photondse}
                                  \includegraphics[width=0.9\columnwidth]{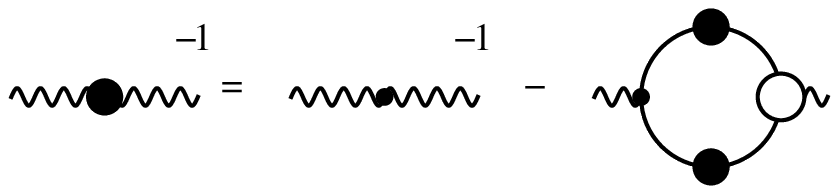}}
	        \end{center}
                \caption{The diagrammatic representation of the Dyson--Schwinger equations
                         for the fermion (a) and gauge boson propagator (b).
                         Wiggly lines denote photon propagators, straight lines 
                         fermion propagators. 
                         A blob denotes a dressed propagator or vertex, whereas a 
                         dot stands for a bare fermion-photon vertex.}
\end{figure}

It enters the Lagrangian in the form of:

\begin{align}
\mathcal{L}^{aniso}=\frac{N_f}{2}\sum_{j=1,2}
                         \bar{\Psi}_{j}
                         \left\{ \sum_{\mu=0}^{2}\gamma_{\nu}\sqrt{g}_{j,\nu\mu}
                                 \left(\partial_{\mu}+\mbox{i}\;a_{\mu}\right)
                         \right\}
                         \Psi_{j}.
\label{eq:lagrangian}
\end{align}
As in isotropic \qed, we can define chiral symmetry using $\gamma_3$ and $\gamma_5$.  
With massless fermions, the Lagrangian therefore has a 
$\;U(2N_f)\;$ ``chiral'' symmetry, which is broken to
$\;SU(N_f) \times SU(N_f) \times U(1) \times U(1)\;$ 
if the fermions become massive. The order parameter for this symmetry
breaking is the chiral condensate which can be determined via the trace 
of the fermion propagator 
$S_{F,j}(\vec{p}) = \int \frac{d^3x}{(2\pi)^3} e^{i\vec{p}\vec{x}} 
\langle \bar{\Psi}_j(\vec{x}) \Psi_j(0) \rangle$.
The corresponding expression for the photon propagator is given by
$D_{\mu\nu}(\vec{p}) = \int \frac{d^3x}{(2\pi)^3} e^{i\vec{p}\vec{x}} 
\langle A_\mu(\vec{x}) A_\nu(0) \rangle$.
The Dyson--Schwinger equations for the fermion and photon propagators 
are diagrammatically represented as shown in 
Fig.~\ref{fig:fermiondse} and Fig.~\ref{fig:photondse}. 
In Euclidean space-time, they are explicitly  given by 
\begin{widetext}
\begin{eqnarray}
S_{F,i}^{-1}(\vec{p}\;)\;&=& S_{0}^{-1}(\,\vec{p}\,)
                          \;\;+\,Z_{1}\,e^{2}\sum_{i=1,2}\int\frac{d^{3}q}{(2\pi)^{3}}
                                        (\sqrt{g}_{i,\mu\alpha}\gamma^{\alpha}\,S_{F,i}(\,\vec{q}\,)\,
                                         \sqrt{g}_{i,\nu\beta}\Gamma^{\beta}(\,\vec{q},\vec{p}\,)\,D_{\mu\nu}(\,\vec{k}\,)),
\label{eq:fermiondse} \\
D_{\mu\nu}^{-1}(\vec{p}\;)  &=& D_{0,\mu\nu}^{-1}(\,\vec{p}\,)
                               -Z_{1}\,e^{2}\,\frac{N_f}{2}\,\sum_{i=1,2}\int\frac{d^{3}q}{(2\pi)^{3}}
                                            \mbox{Tr}\left[
                                                           \sqrt{g}_{i,\mu\alpha}\gamma^{\alpha}\,S_{F,i}(\,\vec{q}\,)\;
                                                           \sqrt{g}_{i,\nu\beta}\Gamma^{\beta}(\,\vec{p},\vec{q}\,)\,S_{F,i}(\,\vec{k}\,))
                                                     \right],
\label{eq:photondse}
\end{eqnarray}
\end{widetext}
with the momentum $\,\vec{k}\,$ defined by the difference $\vec{p}-\vec{q}$ and $i=1,2$. 
Here, the renormalization constant $Z_{1}$ of the fermion-boson vertex,
 $\,\Gamma^{\beta}(\,\vec{p},\vec{q}\;)$, is included.

As we consider an anisotropic space-time 
for fermions, we have to take care of dressing 
each Dirac component separately and also include 
the node index. In order to keep track more easily 
of the general structure of the 
equations, we introduce some shorthands 
denoted by
\begin{equation}
 \overline{p}_{i}^{2}:=p_{\mu}\;g_{i}^{\mu\nu}\;p_{\nu}
\label{eq:barconvention}
\end{equation}
and
\begin{align}
 \widetilde{p}_{\mu,i}:=A_{\mu,i}\left(\vec{p}\;\right) p_{\mu},\mbox{ (no summation convention !), }
\label{eq:tildeconvention}
\end{align}
Where $A_{\mu,i}$ denotes the vectorial 
fermionic dressing function at node $i$.  

The anisotropic expressions for the dressed fermion
and photon propagators then read
\begin{eqnarray}
S_{F,i}^{-1}\left(\vec{p}\;\right)&=&B_{i}\left(\vec{p}\;\right)+\mbox{i}\;\sqrt{g_{i}}^{\mu\nu}\gamma_{\nu}\;\widetilde{p}_{\mu,i},
\label{eq:fermion}
\\
D_{\mu\nu}\left(\vec{p}\;\right)^{^{-1}}&=&p^{2}\left(\delta_{\mu\nu}-\frac{p_{\mu}p_{\nu}}{p^{2}}\right)
                                              +\Pi_{\mu\nu}\left(\vec{p}\;\right),
\label{eq:photon}
\end{eqnarray}
where $B_{i}$ denotes the scalar fermion dressing function at node \emph{i} and 
$\Pi_{\mu\nu}\left(\,\vec{p}\;\right)$ the vacuum polarization of the gauge boson field. 
As the free gauge bosons do not involve fermionic velocities, the bare propagator 
still is the isotropic one.

In order to close the Dyson-Schwinger equations
for the propagators we need to specify the
fermion photon vertex
$\Gamma^\beta$. This vertex satisfies its own
Dyson-Schwinger equation involving a fermion
four-point function of which very little is 
known to date. Strategies to explore the 
fermion-photon interaction in the isotropic
case therefore centered around vertex constructions 
based on higher order perturbative expansions,
multiplicative renormalizability and the Abelian
Ward identity \cite{Curtis:1990zs,Bashir:1999bd,Kizilersu:2009kg}. 
One of the main results of
Ref.~\onlinecite{Fischer:2004nq}, however, is the 
stability of the (isotropic) critical number 
of fermion flavors with respect to the dressing of the 
vertex: the difference between the values
for a bare vertex and much more sophisticated
constructions is smaller than five percent. This
finding somewhat justifies the use of simple
vertex constructions also in the case of
anisotropic space-time, where the complexity of the
DSEs is large and any simplifications are highly
welcome. In this work we therefore employ a vertex
construction which only takes into account the
leading term of the Ball-Chiu vertex \cite{Ball:1980ay} (1BC-vertex), 
generalized to anisotropic space-time. It reads
\begin{equation}
	\Gamma^\beta_i(\vec{p},\vec{q}) = 
	\gamma^\beta \frac{A^\beta_i(\vec{p}) + A^\beta_i(\vec{q})}{2} 
	\label{vertex}
\end{equation}
where no summation convention is used and
$p,q$ are the fermion and anti-fermion momenta 
at the vertex. This vertex is known to be a reasonable 
approximation to the full Ball-Chiu vertex \cite{Maris:1996zg}, which solves 
the Ward-identity exactly. We will use the vertex Eq.~(\ref{vertex})
in the quark DSE and also compare results with simpler
truncation schemes such as the $1/N_f$ approximation or
the Pisarski scheme \cite{Pisarski:1984dj}.

In order to make the photon DSE tractable we will use two different approximations:
the $1/N_f$ approximation and a more sophisticated model based on the results
of Ref.~\onlinecite{Fischer:2004nq}. In the limit of isotropic QED$_3$ there has 
been extensive work in the $1/N_f$ approximation \cite{Appelquist:1988sr,Nash:1989xx}, 
which has been found to agree qualitatively with more elaborate approaches 
\cite{Fischer:2004nq}. In the anisotropic case the leading order $1/N_f$ 
approximation has been pursued in Refs.~\onlinecite{Franz:2002qy,Lee:2002qza}, 
where in addition an expansion in terms of small anisotropies has been performed. 
This is not the case here: we keep the full dependence of the equations on all 
three velocities $(c_s,\vf,\vd)$. As a result, one may expect to get a rough, 
but qualitatively correct picture of the effects of anisotropy close to the 
transition from the chirally symmetric to the chirally broken phase. 

\begin{figure}[b]
		\begin{center}
			\includegraphics[width=0.9\columnwidth]{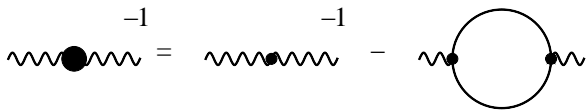}
             \end{center}
                \caption{The diagrammatic representation of the Dyson--Schwinger equation for the vacuumpolarization of
			 gauge bosons in leading order large-$\nf$ expansion.
\label{fig:largenvacuumpolarization} }
\end{figure}

The vacuum polarization in leading order large-$\nf$ expansion is given 
diagrammatically in Fig.~\ref{fig:largenvacuumpolarization}.
The integration can be performed analytically and one obtains \cite{Franz:2002qy,Lee:2002qza}:
\begin{eqnarray}
\Pi^{\mu\nu}\left(\vec{p}\;\right) & =&  \sum_{i}\sqrt{\overline{p}_i^2}
\left(g_{i}^{\mu\nu}-\frac{g_{i}^{\mu\alpha} p_{\alpha}\;g_{i}^{\nu\delta}p_{\delta}}
{\overline{p}_i^2}\right)\,\,
\Pi_i\left(\vec{p}\;\right).\nonumber\\
\Pi_i\left(\vec{p}\;\right) & = & 
\frac{e^{2}\, N_{f}}{16\,v_{F}v_{\Delta}} \frac{1}{\sqrt{\overline{p}_i^2}}
\label{eq:largenvacuumpolarization}
\end{eqnarray}

Despite its merit of simplicity, there is a serious drawback of the leading 
order $1/N_f$-expansion, which makes it desirable to go beyond: it leaves the 
values of the fermion vector dressing functions $A_i = 1$ at all momenta. In 
the isotropic case, this behavior is in marked contrast to the one of more
sophisticated approximations. There, infrared power laws in $A$, the photon 
polarization $\Pi$ and the quark-photon vertex appear, which are characteristic 
for the chirally symmetric phase \cite{Fischer:2004nq}. In addition, in the 
anisotropic case, non-trivial values of the $A$-functions at zero momenta 
serve to define renormalized velocities $(c_s^R,\vf^R,\vd^R)$. It is not clear
whether this is possible within the $1/N_f$-expansion. In order to go beyond
this approximation we therefore utilize the results of Ref.~\onlinecite{Fischer:2004nq}:
in the isotropic case and in the symmetric phase the photon vacuum polarization
is well approximated by the following expression:
\begin{equation}
	\Pi(k) = \frac{e^2 N_f}{8} \left(\frac{1}{k}\frac{k^2}{k^2+e^2} 
	+ \frac{1}{k^{1+2\kappa}}\frac{e^2}{k^2+e^2}\right) \label{isomodel}
\end{equation}
This expression includes the $1/N_f$-limit for $\kappa=0$. In general, however,
this anomalous dimension is non-zero and dependent on the details of
the vertex truncation. In Ref.~\onlinecite{Fischer:2004nq} the value of
$\kappa$ for the vertex truncation Eq.~(\ref{vertex}) at the critical number of
fermion flavors $N_f^c$ has been determined as $\kappa=0.0358$. In principle, 
this number may depend on the fermion velocities in the anisotropic 
case. Nevertheless, due to our current inability to calculate this dependence
we will keep $\kappa$ constant in all calculations. We therefore arrive at the
anisotropic generalization of (\ref{isomodel})  
\begin{equation}
	\Pi_i\left(\vec{p}\;\right) = \frac{e^2 N_f}{16\,v_{F}v_{\Delta}}  
	\left(\frac{1}{\sqrt{\overline{p}_i^2}}\frac{\overline{p}_i^2}{\overline{p}_i^2+e^2} 
	+ \frac{1}{\overline{p}_i^{1+2\kappa}}\frac{e^2}{{\overline{p}_i}^2+e^2}\right) \label{model}
\end{equation}
with  $\overline{p}_{i}^{2}$ defined in Eq.~(\ref{eq:barconvention}).

In general, we wish to emphasize, that great care has to be taken to use
consistent approximations in the photon and fermion DSEs. Either one has to
use the $1/N_f$ approximation in both DSEs, leading to the well known isotropic
result \cite{Appelquist:1988sr} $N_f^c \approx 3.24$ , or one has to 
use the vertex \Eq{vertex} in both DSEs and solve everything self-consistently 
\cite{Fischer:2004nq}, leading to $N_f^c \approx 3.56$. Below we will show
that we reproduce this last result approximately using our model \Eq{model}
for the photon. It is, however, inconsistent and therefore wrong to use the 
$1/N_f$-approximation for the photon polarization 
\Eq{eq:largenvacuumpolarization} and the vertex \Eq{vertex} 
in the quark-DSE. This choice leads to $N_f^c \rightarrow \infty$ and therefore
prohibits any chiral restoration. We also wish to emphasize that to our mind,
meaningful results cannot be extracted from models that violate translation 
invariance of the photon propagator \cite{Bashir:2008fk,Bashir:2009fv}.

Having specified our approximation (\ref{model}) of the photon, we now proceed 
with the fermion-DSE.
In order to extract the scalar and vector dressing function of the fermion
we take appropriate traces of the fermion-DSE and rearrange the equations 
to end up with:

\begin{widetext}
\begin{eqnarray}
%
B_{i}\left(\vec{p}\;\right) &      = & Z_{2} \; e^{2}\int\frac{d^{3}q}{\left(2\pi\right)^{3}}
                                       \,\, \frac{B_{i}\left(\vec{q}\;\right)g_{i}^{\mu\nu}D_{\mu\nu}(\vec{k}\;)}
                                             {B_{i}\left(\vec{q}\;\right)^{2}+(\overline{\widetilde{\vec{q}}}_{i}\;)^{2}}\,\,\,\frac{A_{\mu,i}(\vec{p}) + A_{\mu,i}(\vec{q})}{2},                                 
,
\label{eq:bdse}
                                 \\
%
A_{\mu,i}\left(\vec{p}\;\right) &    = & Z_{2}-\frac{Z_{2}\;e^{2}}{p_{\mu}}
                                        \int\frac{d^{3}q}{\left(2\pi\right)^{3}}
                                       \,\, 
                                            \frac{2\left(\widetilde{q}_{\lambda,i}\;g_{i}^{\lambda\nu}D_{\mu\nu}  (\vec{k}\;)\right)
                                                        -\widetilde{q}_{\mu,i}\;      g_{i}^{\lambda\nu}    D_{\lambda\nu}(\vec{k}\;)}
                                                 {B_{i}\left(\vec{q}\;\right)^{2}+(\;\overline{\widetilde{\vec{q}}}_{i}\;)^{2}}
                                        \,\, 
\frac{A_{\mu,i}(\vec{p}) + A_{\mu,i}(\vec{q})}{2},                                 
\end{eqnarray}
\end{widetext}
where again we have no summation for the external index $\mu$. Furthermore we have used the
Ward-identity $Z_1=Z_2$ between the fermion wave function renormalization factor $Z_2$ and the
one for the vertex, $Z_1$. 
The form of the ``\emph{metric}'' leads to several symmetries that 
will simplify the evaluation of the DSEs. 
In general, we find the relations:
%
\begin{eqnarray}
B_{1}    \left(\,p_{0},p_{1},p_{2}\right) & = & B_{2}    \left(\,p_{0},p_{2},p_{1}\right),   \label{eq:symmetries of b}\\
A_{\mu,1}\left(\,p_{0},p_{1},p_{2}\right) & = & A_{\mu,2}\left(\,p_{0},p_{2},p_{1}\right),   \label{eq:symmetries of a}\\
\Pi^{11} \left(\,p_{0},p_{1},p_{2}\right) & = & \Pi^{22} \left(\,p_{0},p_{2},p_{1}\right),   \label{eq:symmetries of pi}\\
\Pi^{10} \left(\,p_{0},p_{1},p_{2}\right) & = & \Pi^{20} \left(\,p_{0},p_{2},p_{1}\right).
\end{eqnarray}
One then ends end up with eight anisotropic equations that need to be 
solved self-consistently, in comparison to only three equations for an 
isotropic space-time. This increase in the number of equations results from 
the component-wise dressing of the fermion momentum and from the anisotropic
vacuum polarization of the gauge bosons. In our case using the $1/N_f$-approximation
or our model (\ref{model}) in the photon sector we remain with four equations.
These reduce to two in the isotropic limit.

\subsection{The DSEs on a torus \label{subsec:dsetorus}}

Although our approximations of the photon serve to simplify the system of DSEs
dramatically, the numerical procedure to solve the remaining equations for 
$B_1,A_{0,1},A_{1,1},A_{2,1}$ is still considerable due to the presence of the
anisotropy. In particular, the usual transformation to hyper-spherical coordinates
and exploitation of symmetries therein is of course no longer feasible. One possibility to circumvent this problem is to
change the base manifold of the system from three-dimensional Euclidean space to that
of a three-torus. In Lorenz-covariant systems such a treatment has been explored
extensively in the context of QCD \cite{Fischer:2002eq,Fischer:2005nf} and also for 
QED$_3$ \cite{Goecke:2008zh}. Here we generalize this treatment to the anisotropic
case. In fact, the formulation of DSEs on a torus is ideally suited for
this endeavor,
since a formulation in Cartesian instead of hyper-spherical coordinates is most natural. 

On a compact manifold, the photon and fermion fields have to obey appropriate
boundary conditions in the time direction. These have to be periodic for
the photon fields and anti-periodic for the fermions. For computational reasons 
it is highly advantageous, though not necessary, to choose the same conditions 
in the spatial directions. We choose the box to be of equal length in all directions,
$L_1=L_2=L_3\equiv L$, and denote the corresponding volume $V=L^3$. Together with the 
boundary conditions this leads to a discretization of momentum space. Thus all 
momentum integrals appearing in the Dyson-Schwinger equations are replaced by sums 
over Matsubara modes. For the fermions this amounts to
\begin{equation}
\int \frac{d^3q}{(2 \pi)^3} \:(\cdots) \:\:  \longrightarrow \:\:\frac{1}{L^3}
\sum_{n_1,n_2,n_3} \:(\cdots) \,, 
\end{equation}
counting momenta ${\bf q}_{\bf n} = \sum_{i=1..3} (2\pi/L)(n_i+1/2) \hat{e}_i$, where
$\hat{e}_i$ are Cartesian unit vectors in Euclidean momentum space. For the
photon with periodic boundary conditions the momentum counting goes like 
${\bf q}_{\bf n} = \sum_{i=1..3} (2\pi/L)(n_i) \hat{e}_i$.
The resulting system of equations can be solved numerically along the lines
described in Ref.~\onlinecite{Goecke:2008zh}. 

Of course, working on a torus with a finite volume introduces artifacts. These
have been studied in some detail in Ref.~\onlinecite{Goecke:2008zh}, and
found to be sizable, see Fig.~7 of this reference. In addition, nonlinear 
volume effects have been found for very large volumes in 
Ref.~\onlinecite{Gusynin:2003ww}. The reason for these large effects is the
large separation of scales in QED$_3$. Besides the natural scale $e^2=1$ of the
coupling with dimension one, there is the scale of dynamical mass generation which
even well in the broken phase is at least two orders of magnitude below the
natural scale. This is a situation totally different to, \emph{e.g.}, QCD where both
scales are of the same order. In the isotopic theory, the large volume effects 
in QED$_3$ explain why lattice simulations 
\cite{Hands:2002dv,Hands:2004bh,Strouthos:2008hs}
find critical numbers of fermion flavors of the order of $N_f^c \approx 1.5$, while
continuum calculations in the DSE framework \cite{Fischer:2004nq} find $N_f^c \approx 3.5$.
The DSE calculations on the torus explained these two results as
a consequence of the finite lattice volume \cite{Goecke:2008zh}.
We will briefly address the corresponding volume effects in the anisotropic case in section 
\ref{subsec:volumestudy}. We find indications, that the size of the volume effects 
is hardly affected by changes of the fermion velocities, at least in the range 
relevant for our study. Thus while the absolute number for $N_f^c$ on the torus 
needs to be corrected, its variation with $(c_s,\vf,\vd)$ is nevertheless 
meaningful. We therefore adopt a similar strategy as the authors of 
Ref.~\onlinecite{Concha:2009zj} and concentrate on the variations rather 
than the absolute value of $N_f^c$.

\section{Numerical results \label{results}}

In this section we will discuss our numerical results for three different
approximation schemes. For the fermion DSE these are 
(i) the Pisarski scheme \cite{Pisarski:1984dj}, 
(ii) the leading order $1/N_f$ expansion and (iii) our most elaborate scheme
using the vertex Eq.~(\ref{vertex}) and the photon model \Eq{model}. 
Before we present our results for the isotropic and anisotropic cases
at finite volume, we briefly discuss the corresponding infinite volume 
results in the isotropic case. In general, besides the chiral condensate 
another suitable order parameter for 
chiral symmetry breaking is the value of the scalar fermion dressing function
at any point in momentum space; here we use $B(\vec{p})$ at the smallest
momentum available.   

\subsection{The critical $\mbox{N}_{f}^{c}$ in the isotropic case}

In the Pisarski scheme, analysed in detail in Ref.~\onlinecite{Pisarski:1984dj},
the fermion-photon vertex is set to the bare one and the scalar dressing
function of the fermion is set to a constant, $B(\vec{p})=B(0)=m$. This rough 
approximation scheme is well suited to analytically demonstrate that chiral
symmetry is broken in QED$_3$. As a result one obtains
\begin{equation}
m = B(0) = c\,\alpha\, e^{-\pi^2 N_f/8},
\end{equation}
where $c$ is a positive constant and $\alpha=e^2 \, N_f$. Clearly, chiral
symmetry is broken and remains broken for any value of $N_f$. Thus 
$N_f^c \rightarrow \infty$, which is in contrast to the findings in more 
elaborate approximation schemes. Therefore, although the Pisarski scheme 
served as a nice tool for demonstrating broken chiral symmetry in the first 
place, it is not well suited to extract information on chiral symmetry restoration. 
We will see below that this remains true in the anisotropic case.

\begin{figure}[t]
\begin{center}
\includegraphics[width=0.9\columnwidth]{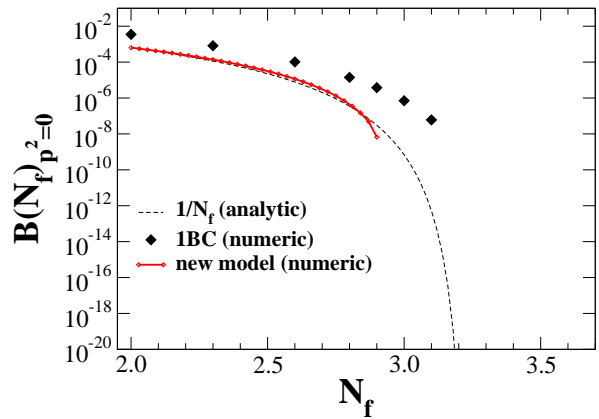}
\end{center}
      \caption{The order parameter for chiral symmetry breaking, $B(0)$, plotted
      as a function of the number of fermion flavours $N_f$ in the isotropic case
      $1=c_s=\vf=\vd$. Shown are results
      for the selfconsistent leading order $1/N_f$-approximation and the 1BC-vertex,
      \Eq{vertex}, compared with the results for our model \Eq{model} for the
      photon together with the vertex \Eq{vertex} in the fermion-DSE.\label{newmodel_isotropic}}
\end{figure}

Let us now focus on the other two approximation schemes (ii) and (iii). In 
the infinite volume limit the dependence of $B(0)$ on $N_f$ has been determined in
Ref.~\onlinecite{Fischer:2004nq} together with results for more elaborate vertex truncations.
Here we combine the result for our photon model \Eq{model} together with the vertex
approximation Eq.~(\ref{vertex}). In Fig.~\ref{newmodel_isotropic},
we display our results. For all three truncations we clearly find an exponential
decrease of the order parameter with $N_f$, a situation described as
Miransky-scaling \cite{Miransky:1988gk} in the literature. (Note that universal power 
law corrections to Miransky scaling have been discussed in Ref.~\onlinecite{Braun:2010qs}
and Refs. therein. For QED$_3$ the quantitative aspects of these corrections
have not yet been studied.) The critical number $N_f^c$ has been determined
analytically for the $1/N_f$ and the self-consistent 1BC-vertex truncations:
$N_f^{c,1/N_f} \approx 3.24$ and $N_f^{c,1BC} \approx 3.56$. The corresponding 
value for our model $N_f^{c,model}$ is a little lower than the one for the 
$1/N_f$-approximation, as can be inferred from Fig.~\ref{newmodel_isotropic}. 

The reason why $N_f^{c,model} \ne N_f^{c,1BC}$ is the following: the photon in 
the self-consistent 1BC-approximation is a constant in the infrared for 
$N_f < N^c_f$ and develops an infrared power law for $N_f > N^c_f$. In our model,
the power law is already present for $N_f < N^c_f$, which results in missing 
infrared strength and therefore explains why $N_f^{c,model} < N_f^{c,1BC}$.
This is, however, only a small quantitative effect which is irrelevant for the
qualitative study presented in this work.

\subsection{The critical $\mbox{N}_{f}^{c}$ in the anisotropic case \label{subseq:phasediagrams}}

In this section we discuss the general anisotropic case with $c_s=1$ 
and $(\vf,\vd)$ varied. We first investigate the Pisarski approximation,
which has been used in Ref.~\onlinecite{Concha:2009zj} to argue for a couple 
of general points. Again, similar to the isotropic case, we can determine
the behavior of the order parameter analytically, at least for $\vf=\vd$.
The details are reported in appendix \ref{app:pisarski}. We find
%
\begin{eqnarray}
m &=& B(0) = c\,\alpha\, \exp\left[\left(-\pi^2 N_f \sqrt{\vf^2-1}\right)/ \right. 
             \label{eq:aniso_pisarski}\\
&&\left.\left(2\sqrt{\vf^2-1}+2(2+\vf^2)\arctan(\sqrt{\vf^2-1})\right)\right],\nonumber
\end{eqnarray}
%
which is different to the corresponding expression reported in 
Ref.~\onlinecite{Concha:2009zj}. Our result has the correct isotropic 
limit and it generalises the finding from the isotropic case:
in the Pisarski approximation (and for $d=3$ dimensions) chiral symmetry 
is always broken, i.e. $N_f^c \rightarrow \infty$. We therefore find 
this approximation not to be suited in general to extract quantitative 
information on the variation of $N_f^c$ with $(c_s,\vf,\vd)$. 

\begin{figure}[t!]
\begin{center}
\includegraphics[width=0.9\columnwidth]{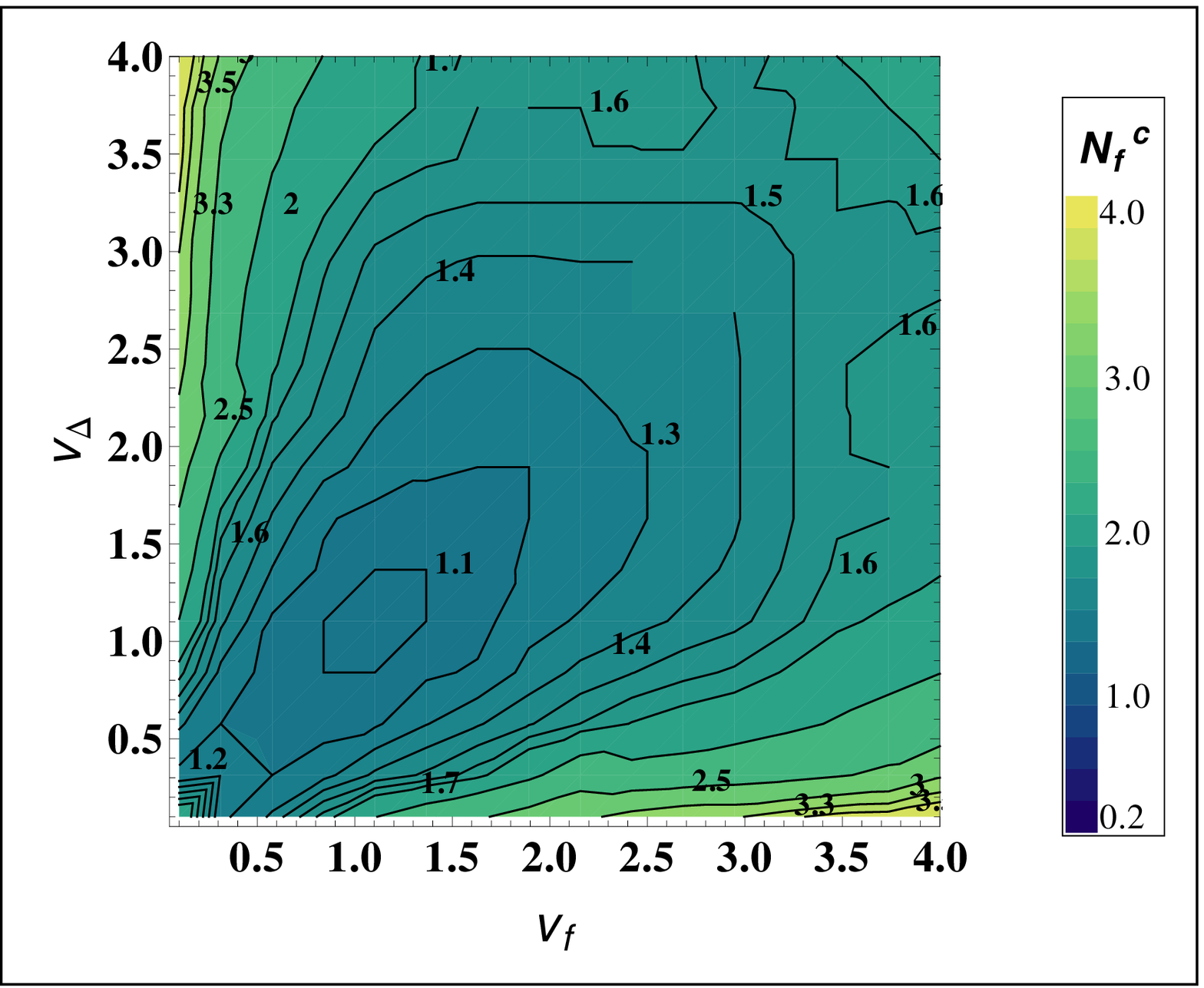}
\end{center}
      \caption{Critical number of fermion flavours $N_f^c$ determined in the 
      $1/N_f$-approximation and plotted as a function of the fermion velocities $(\vf,\vd)$.      
      \label{res:1/N}}
\begin{center}
\includegraphics[width=0.9\columnwidth]{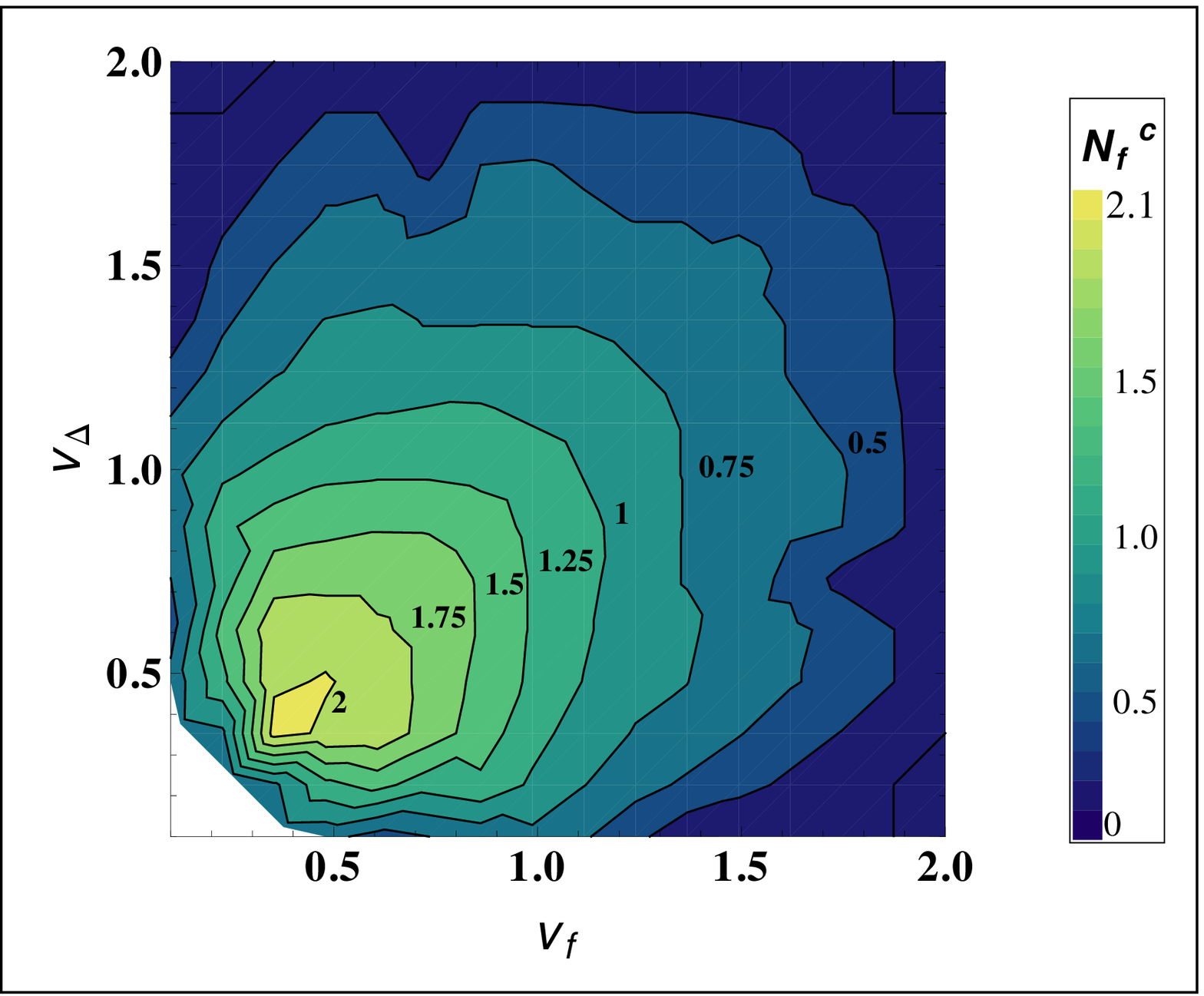}
\end{center}
      \caption{Critical number of fermion flavours $N_f^c$ determined from the
      photon model \Eq{model} and the fermion-photon vertex \Eq{vertex} 
      plotted as a function of the fermion velocities $(\vf,\vd)$.
      \label{res:model}}
\end{figure}

We now proceed with our two more elaborate approximations. In Figs.~\ref{res:1/N}
and \ref{res:model} we show our results on a torus with $e^2=1$, $L=600/e^2$ and
$20^3$ momentum points. The values of $N_f^c$ are represented by contour lines and 
a corresponding color code. Note that the `edges' in the contour lines are due to 
the sparseness of our grid in some regions; in general we expect smooth lines
in the $\vf\mbox{-}\vd$-space. For $\vf=\vd=1$ we can read off the corresponding
values for $N_f^c$ in the isotropic case. We find $N_f^{c,1/N_f} \approx 1.0$
and $N_f^{c,model} \approx 1.1$. Clearly, these values are affected by large 
finite volume effects, as already discussed above. When we compare these with 
the corresponding values in the continuum calculation we find that the latter 
are roughly a factor of three larger. This has to be kept in mind when discussing
Figs.~\ref{res:1/N} and \ref{res:model}.

For non-trivial values of
$\vf$ and $\vd$ we find sizable deviations of $N_f^c$ from the isotropic
case. We also observe that the two truncation schemes behave very differently.
In general the $1/N_f$-approximation shows an increase of the critical
number of fermions with respect to deviations from the isotropic case in all 
directions. This finding is in agreement with early expectations 
\cite{Lee:2002qza} from Dyson-Schwinger equations, performed in the 
$1/N_f$-expansion and furthermore in an expansion of small anisotropies. 
Thus what we have shown here is that the results of Ref.~\onlinecite{Lee:2002qza}
survive also for large anisotropies. There remains, however, the question
of the reliability of the $1/N_f$-expansion. We have already discussed above, 
that the leading order $1/N_f$-expansion contains the approximation $A_i = 1$,
which is certainly not valid in general. Our modified approximation scheme
\Eq{vertex} and \Eq{model} avoids this problem. Indeed, our results for $N_f^c$
in this model show a different behavior. Here we find a
maximum of $N_f^c$ for small values of $(\vf,\vd)$ at or around
$(\vf,\vd) = (0.4,0.4)$ with {\it decreasing} behavior for smaller and 
larger velocities. In the region where both velocities are smaller
than $c_s=1$ the critical number of fermions never goes to zero. This is
different for large velocities: in fact if one of the velocities is larger 
than a certain critical value of the order of $\vf,\vd \approx 2$, 
the system is always in the chirally symmetric phase. Certainly the precise number,
where this happens, is expected to be severely affected by volume effects. Nevertheless,
this qualitative behavior is in good agreement with the corresponding
evaluation performed in Ref.~\onlinecite{Concha:2009zj}, where the criterion of
the strength of one-photon exchange between fermions was used. Our results therefore 
corroborate their findings. (It is perhaps amusing to note that in 
Ref.~\onlinecite{Concha:2009zj} the statement was made that it is `almost
impossible' to analyze mass generation in the presence of anisotropies
using Dyson-Schwinger equations. Naturally, in the light of this work we
do not agree with this statement.) As argued in detail in 
Ref.~\onlinecite{Concha:2009zj}, the picture indicated by Fig.~\ref{res:model} 
is also supported by lattice calculations \cite{Hands:2004ex,Thomas:2006bj}.
In the light of this agreement we tend
to the statement that the $1/N_f$-approximation picture of Fig.~\ref{res:1/N} is
misleading and has to be discarded.

\subsection{Volume effects and the case $\vf / \vd \sim 10$
\label{subsec:volumestudy}}

\begin{figure}[t]
\begin{center}
\includegraphics[width=0.9\columnwidth]{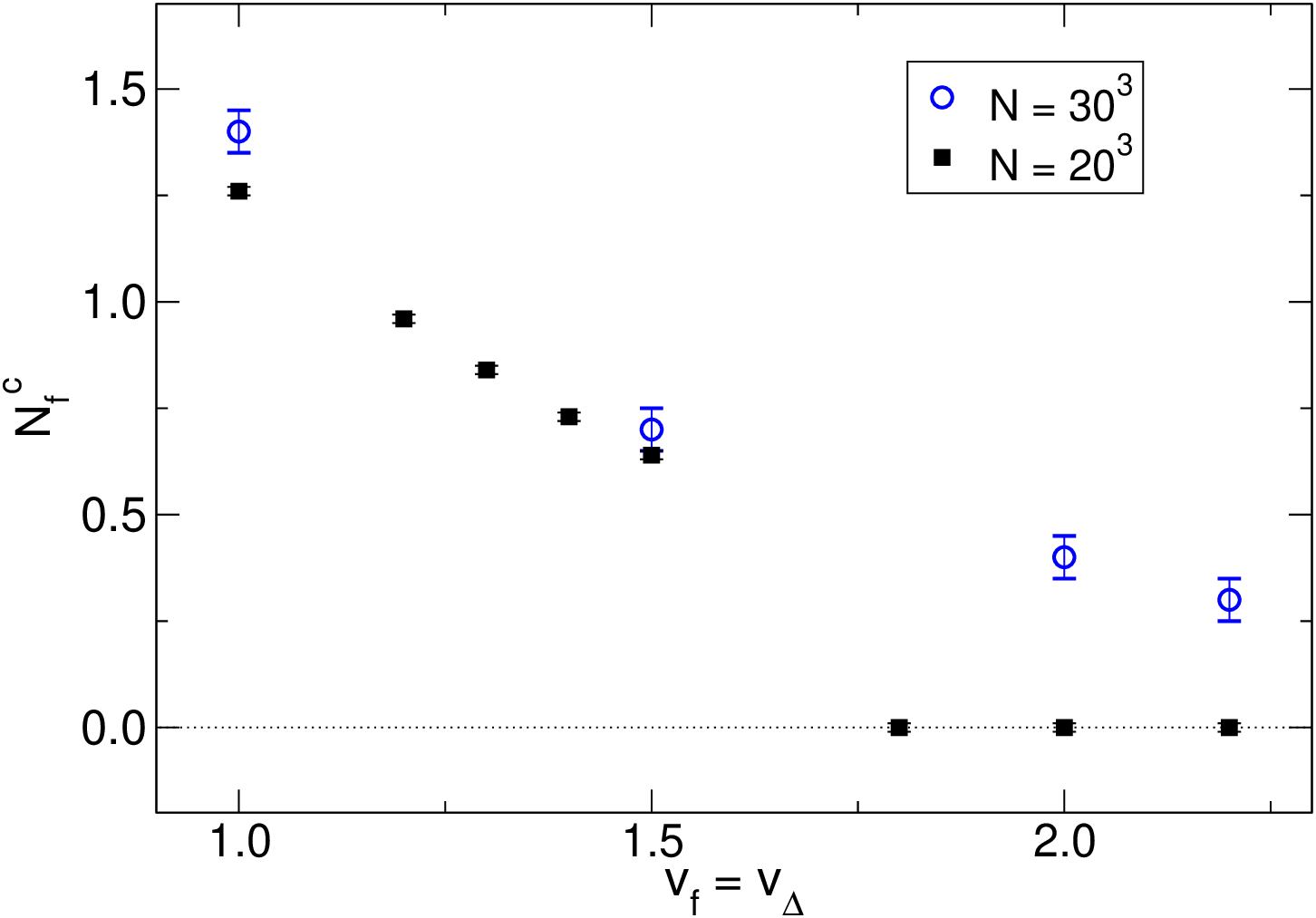}
      \caption{Four selected points on the $\vf = \vd$-axis determined with 
      two different tori with $N=20^3$ and $N=30^3$ corresponding to box lengths 
      of $L e^2 = 600$ and $L e^2 = 900$.
      \label{volume}}
\includegraphics[width=0.9\columnwidth]{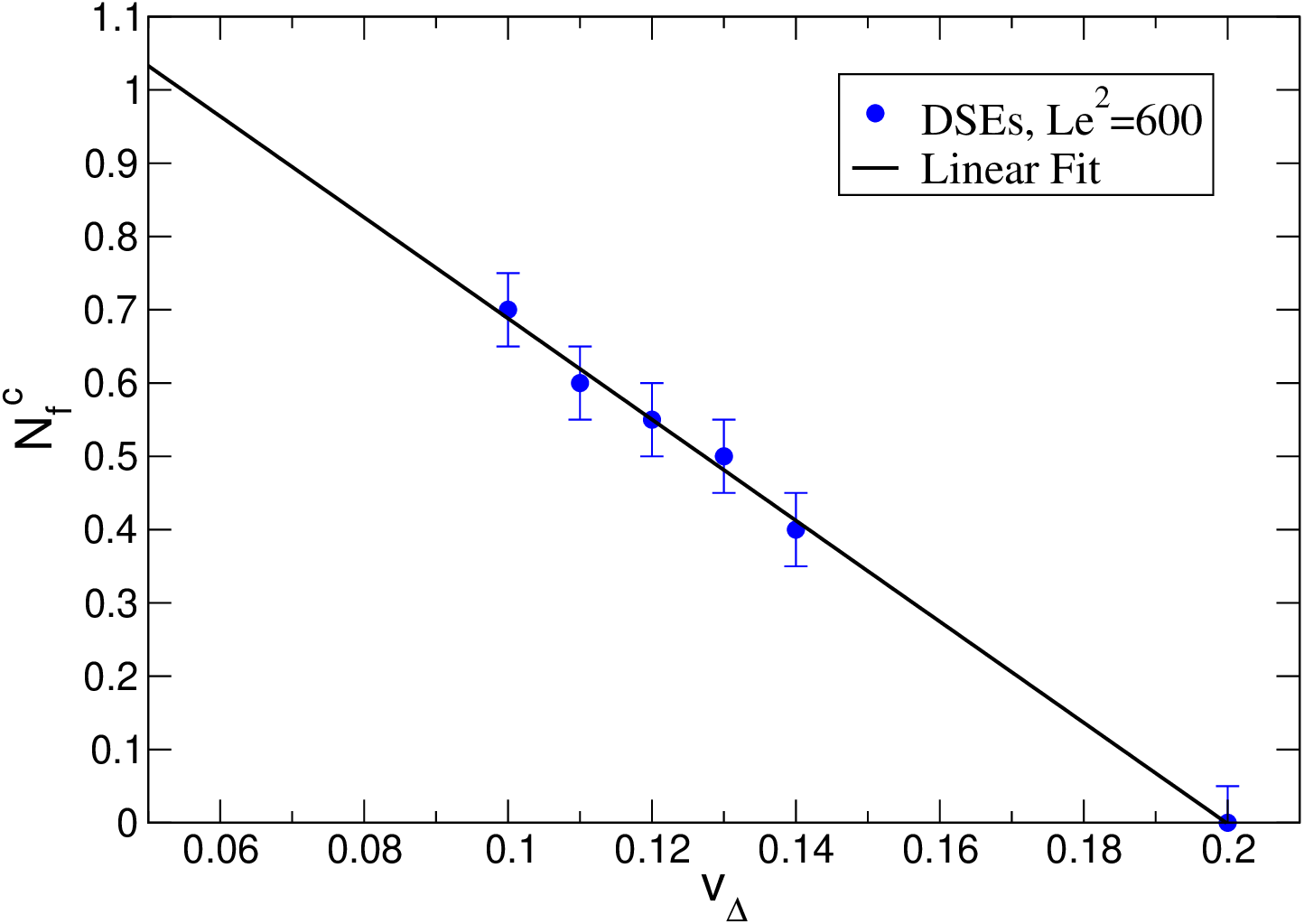}
      \caption{Selected results for the critical number of fermion
      flavors at $(\vf,\vd)=(10\vd,\vd)$ from a torus with $L e^2 = 600$.
      The error bars correspond to our step size when searching for $N_f^c$.
      \label{physical}}
\end{center}
\end{figure}

Let us now briefly come back to the volume effects associated with our 
calculation on the torus. In Ref.~\onlinecite{Goecke:2008zh} these effects
have been analyzed in the isotropic case and tori as large as $N = 512^3$
have been used. In the anisotropic case the numerical effort to determine
$N_f^c$ is much larger than in the isotropic case and the $N=20^3$ calculations
reported above are already demanding in terms of CPU time. A detailed and
concise study of the volume effects using large tori is therefore beyond the 
scope of the present work. Nevertheless we wish to demonstrate, that the assumption 
used above, namely that the size of the volume effects do not too strongly depend
on the size of the anisotropy, may be correct. To this end we show the results for 
four selected points on the $\vf = \vd$-axis for $N=30^3$ in Fig.~\ref{volume}.
Clearly, the larger volume results in a larger value for $N_f^c$, as visible from
the plot. Comparing the volume effects at the isotropic point with the value
at $\vf=1.5$ we find quantitative differences, but certainly not too different 
to invalidate our assumption from the last section on a qualitative level. 
Finally we also observe, that the line of demarcation from $N_f^c \ne 0$ to 
$N_f^c=0$ moves strongly with volume. Clearly, further studies are necessary to 
pin down the size of the volume effects to such
an extent, that extrapolations to the infinite volume limit along the lines
of Ref.~\onlinecite{Goecke:2008zh} are possible. To forego the problem
of finite volume effects, one may also attempt to solve the
equations directly in the infinite volume continuum limit. We will report on first
results in this direction elsewhere. 

When comparing with experiment, values of $(\vf,\vd)$ of the order of
$\lambda = \vf / \vd \sim 10$ may be of particular interest; Ref.~\onlinecite{Chiao:2000}
reports $\lambda = 14$ for optimally doped $\mbox{YBa}_2\mbox{Cu}_3\mbox{O}_7$
and $\lambda=19$ for $\mbox{Bi}_2\mbox{Sr}_2\mbox{CaCu}_2\mbox{O}_8$ compounds.
In addition, it is important to know the relation of $(\vf,\vd)$ to the speed 
$c_s$ related with the gauge boson. Here, we recall that $c_s$
is not the speed of light in the cuprate compound, but the speed characteristic for
the vortex-antivortex excitations that are represented by the gauge field in the
effective theory. It is thus possible and even anticipated that $\vf > c_s$,
as discussed in Ref.~\onlinecite{Concha:2009zj}. To our knowledge, the precise relation 
of $\vf$ and $c_s$ is not yet determined from experiment. 
We therefore show results for $(10\vd,\vd)$ in a typical range accessible
in our calculation in Fig.~\ref{physical}. The filled circles in the plots are our 
numerical results, whereas the black line is a linear fit to the data points. Clearly, 
the dependence of $N_f^c$ on $\vd$ is well approximated by this fit. Whether this linear
dependence remains intact in the infinite volume limit should be studied in future work.

\section{Summary and conclusions} \label{summary}

In this work we have proposed an approximation for the Dyson-Schwinger
equations of QED$_3$, which is well suited to investigate the effects of
anisotropies on the critical number of fermion flavors for chiral
symmetry restoration. This problem is important for the description
of high $T_c$ cuprate superconductors using QED$_3$ as an effective theory. Our results
support corresponding ones from lattice calculations \cite{Hands:2004ex,Thomas:2006bj} 
and continuum studies of the fermion-photon interaction strength \cite{Concha:2009zj}.
In our best approximation scheme we find a maximum of the critical number of fermion 
flavors $N_f^c$ around $\vf=\vd \approx 0.5$. From this point, $N_f^c$ is decreasing 
in all directions in the $(\vf,\vd)$-plane. 
Further studies, however, are necessary to pin down the 
quantitative aspects of this behavior. Our results are still preliminary in at least 
three aspects. They are affected by large volume effects, which have been studied
in detail only at the isotropic velocity point \cite{Goecke:2008zh}. Our calculation
is certainly affected by truncation errors in the fermion-photon vertex, although we 
have argued from comparison with more systematic calculations at the isotropic 
point \cite{Fischer:2004nq}, that these effects are on a moderate level of the order 
of twenty to thirty percent. Finally, our treatment of the anisotropic photon DSE 
remains preliminary and a more self-consistent calculation is certainly desirable. 
Nevertheless, we believe that the qualitative aspects of our results are correct
and provide a further step forward to understand the connection between QED$_3$
and experimental studies of high $T_c$ superconductors.

\section*{Acknowledgement}
This work was supported by the Helmholtz-University Young Investigator Grant 
No.~VH-NG-332 and by the Deutsche Forschungsgemeinschaft through SFB 634. 

\begin{appendix}
\section{Anisotropic $N_f^c$ in the Pisarski approximation}\label{app:pisarski}

In this appendix we give some details to the dependence of the fermion mass
on the number of flavors in the Pisarski approximation.
For $\vf = \vd $, we insert the large-$N_f$ vacuumpolarization, 
Eq.~(\ref{eq:largenvacuumpolarization}), into the scalar fermion
dressing function given implicitly in Eq.~(\ref{eq:fermiondse}). For equal 
fermionic velocities, we make use of the inherent symmetry and switch to 
cylindrical coordinates. With the abbreviations 
\begin{widetext}
\begin{eqnarray*}
 \mathcal{A} & = & 2 \vf^6 z^4 (\sqrt{r^2 + z^2} \alpha^2 +
                   16 z^2 (4 \sqrt{r^2 + z^2} + \alpha))+
                   8 r^6 (8 \sqrt{r^2 + z^2} + \alpha +
                   \vf^2 (8 \sqrt{r^2 +z^2} + 3\alpha))\\               
 \mathcal{B} & = & r^2 \vf^2 z^2 ((1 + \vf^2)^2 \sqrt{r^2 + z^2} 
                   \alpha^2 + 8 z^2 
                   (8 \sqrt{r^2 + z^2} + 24 \vf^2 \sqrt{r^2 + z^2}                                         
                   +16 \vf^4 \sqrt{r^2 + z^2} + 
                   (2 + 5 (\vf^2 + \vf^4))\alpha))\\
\mathcal{C}  & = & 2 r^4 (\vf^2 \sqrt{r^2 + z^2}\alpha^2 + 
                   4 z^2 (8 \sqrt{r^2 + z^2} + \alpha                                                   
                   +\vf^2 (16 \sqrt{r^2 + z^2}  + 24 \vf^2 \sqrt{r^2 + z^2} 
                   +(5 + 5 \vf^2 + \vf^4)\alpha)))\\
\mathcal{D}  & = & N_f (r^2 + z^2) (r^2 + 
                   \vf^2 z^2)^2 (8 \sqrt{r^2 + z^2} + \alpha)\\
\mathcal{E}  & = & 64 r^4 + r^2 (64 (1 + \vf^2) z^2 + 8 \sqrt{r^2 + z^2} \alpha
                   + \vf^2 \alpha (8 \sqrt{r^2 + z^2} + \alpha))+
                     \vf^2 z^2 (64 z^2 + \alpha (16 \sqrt{r^2 + z^2} + \alpha))\\
\end{eqnarray*}
\end{widetext}
we arrive at 
\begin{equation}
   B(\vec{p})     =  \frac{1}{(2 \pi)^3}
                   \int_0^{2 \pi}\int_0^{\infty}\int_0^{\infty}d\varphi dr dz 
                   \frac{ 16\, B(\vec{p}) \,r \alpha \left(\mathcal{A} + \mathcal{B} + \mathcal{C}
                                        \right)}
                        {\mathcal{D} \mathcal{E}}.
\end{equation}
At this point we perform Pisarski's approximation \cite{Pisarski:1984dj} and set 
the scalar fermion dressing function $B(\vec{p})$ equal to a constant 
$B(\vec{p}) = m$ for all momenta in the range $\alpha \gg p \gg m$. Other
momentum contributions are neglected in the integration. Furthermore, 
since $r < \alpha$ and $q < \alpha$ we only keep terms with the largest 
power of $\alpha$ in the numerator and denominator of the integrand. 
This leads to the more compact equation
\begin{widetext}
\begin{equation}
 m =\frac{1}{(2 \pi)^3}
           \int_0^{2 \pi}d\varphi \int_m^{\alpha} dr \int_m^{\alpha} dz
           \frac{16 m r \left( 2r^4 + r^2(1+\vf^2)^2 z^2 + 2\vf^4 z^4
                       \right)}
                {N_f (r^2 + z^2)^{\frac{3}{2}} (r^2 + \vf^2 z^2)^2}.
\end{equation}
As a next step, we perform the $z$ integral and again neglect the terms not proportional to 
$\alpha$. We end up with
\begin{equation}
 m =\frac{1}{(2 \pi)^3}
    \int_m^{2 \pi}d\varphi \int_0^{\alpha} dr 
    \frac{8 m \left(\sqrt{\vf^2 -1} \alpha + (2 + \alpha) \arctan [\sqrt{\vf^2 - 1}]
              \right)}
         {N_f r \alpha \sqrt{\vf^2 -1}}.
\end{equation}
We further perform the $r$ and $\varphi$ integral and use one last time that 
$\alpha$ is the largest scale in the problem. We finally arrive at
\begin{equation}
  m = \frac{2 m \left( \sqrt{ 1 + \vf^2} + (2 + \vf^2) \arctan(\sqrt{1 + vf^2})
                \right)}
                {\pi^2 N_f \sqrt{ 1 + \vf^2}} \ln\left(\frac{\alpha}{m}\right),
\end{equation}
which leads to the result given in the main text,
\Eq{eq:aniso_pisarski}:
\begin{equation}
m = B(0) = c\,\alpha\, \exp\left[\left(-\pi^2 N_f \sqrt{\vf^2-1}\right)/ \right. 
           \left.\left(2\sqrt{\vf^2-1}+2(2+\vf^2)\arctan(\sqrt{\vf^2-1})\right)\right].\nonumber
\end{equation}
\end{widetext}

\end{appendix}


\end{document}